\documentclass[a4paper,fleqn]{cas-sc}
\usepackage{bm}
\usepackage{subfigure}

\usepackage[numbers, sort&compress]{natbib}

\def\tsc#1{\csdef{#1}{\textsc{\lowercase{#1}}\xspace}}
\tsc{WGM}
\tsc{QE}
\tsc{EP}
\tsc{PMS}
\tsc{BEC}
\tsc{DE}

\begin{document}

\let\WriteBookmarks\relax
\def\floatpagepagefraction{1}
\def\textpagefraction{.001}

\shorttitle{...}

\shortauthors{Klimanova O. et~al.}

\title [mode = title]{Accelerating global search of adsorbate molecule position using machine-learning interatomic potentials with active learning}                      



%
\author[1]{Olga Klimanova}[orcid=0009-0001-7685-4876]





\affiliation[1]{organization={Skolkovo Institute of Science and Technology},
    country={Russian Federation}}

\author[1,2]{Nikita Rybin}[orcid=0000-0001-7053-5295]

\author[1,2]{Alexander Shapeev}[orcid=0000-0002-7497-5594]
\cormark[1]


\affiliation[2]{organization={Digital Materials LLC},
    country={Russian Federation}}

\cortext[cor1]{Corresponding author}



\begin{abstract}
We present an algorithm for accelerating the search of molecule's adsorption site based on global optimization of surface adsorbate geometries. Our approach uses a machine-learning interatomic potential (moment tensor potential) to approximate the potential energy surface and an active learning algorithm for the automatic construction of an optimal training dataset. To validate our methodology, we compare the results across various well-known catalytic systems with surfaces of different crystallographic orientations and adsorbate geometries, including CO/Pd(111), NO/Pd(100), NH$_3$/Cu(100), C$_6$H$_6$/Ag(111), and CH$_2$CO/Rh(211). In the all cases, we observed an agreement of our results with the literature.
\end{abstract}



\begin{keywords}
Adsorption energy \sep Active learning \sep Interatomic potentials \sep Moment Tensor Potentials
\end{keywords}

\maketitle
\section{Introduction}

Catalysis plays an important role in industrial chemistry, affecting sectors such as pharmaceuticals \cite{pharma}, fertilizers in general \cite{ammonia} and syngas conversion \cite{syngas}.
Nowadays, the development of novel, advanced catalysts requires a deep understanding of the catalytic reactions processes at the atomistic level.
Experimentation is often the conventional approach to attain this understanding.
However, this approach can be costly.
Therefore, there is growing attention to the application of computational methods to facilitate this understanding~\cite{cat_1, cat_2, cat_3}.
As an example, the recent Open Catalyst Project~\cite{open_catalyst} aimed to use artificial intelligence to identify new catalysts for renewable energy storage applications.
In this context, the authors presented an overview of the challenges associated with identifying suitable electrocatalysts using quantum mechanical simulations at the level of density-functional theory and machine learning models trained on Open Catalyst datasets \cite{OC20, OC22}.

One of the key descriptors in the field of computational catalysis is the adsorption energy \cite{descriptor_1, descriptor_2, descriptor_3}, which is calculated by determining the configuration of the adsorbate and the surface that minimizes the overall energy of the structure.
In general, determining the adsorption energy presents a number of complexities due to the fact that there can be multiple potential binding sites for an adsorbate on the surface, and for each binding site, there can be multiple ways to orient the adsorbate.
The most common approach for calculating the adsorption energy is based on density functional theory (DFT).
As the computational cost of DFT calculations increases with the number of atoms and electrons in the system, there is a clear motivation to develop tools that decrease this cost.
Consequently, the number of studies dedicated to low-cost methods has increased each year, with the latest including artificial intelligence and machine learning techniques \cite{GNN, AdsorbML, GAP_ads}.
In this work, we propose the use of a moment tensor potential (MTP), as proposed by \citet{Shapeev2016}, as a means of accelerating global optimization of surface adsorbate geometries.
MTP has already been employed successfully  in the exploration of new binary and ternary alloys \cite{GUBAEV2019148}, molecular crystals \cite{mol_crystals}, and boron allotropes \cite{boron}.
In our work, we utilize MTP as a model for interatomic interactions to speed up the optimization of the adsorbate-surface configuration.

As will be demonstrated here, the MTP facilitates fast and precise relaxation of molecules on surfaces, with subsequent energy estimation at a level of accuracy comparable to \textit{ab initio} calculations.
The primary enhancement in efficiency we obtain is as follows: rather than performing all optimization steps within the DFT framework, we relax structures using MTP until the forces acting on atoms vanish.
This approach leads to a substantial reduction in the necessary computational resources.
In the final stage, the most energetically favorable structures are identified, and single-point calculations are performed within the DFT framework to evaluate the results and select the most energetically favorable structure. 

Here, the construction of MTP was done using an active learning approach based on the so-called maxvol algorithm \cite{Podryabinkin2017}, which allows us to select structures based on a well-established selection process and construct an accurate machine-learned potential using a minimal amount of data.
We conducted several experiments to demonstrate MTP's effectiveness and reliability in addressing adsorption site search for various well-known systems in catalysis application: CO/Pd(111), NO/Pd(100), NH$_3$/Cu(100), C$_6$H$_6$/Ag(111), and CH$_2$CO/Rh(211).
We intended to provide adsorption energy calculations for a range of molecule geometries and facets, including low-index (100) and (111) as well as stepped (211). 
The findings of our study demonstrate the capability for utilizing MTP to make accurate predictions of molecules position on the surface.
We reserve the exploration of more complicated systems, such as adsorption on nanoparticles.

The present manuscript is organized as follows.
In Section \ref{Method}, we present our methods.
In particular, the workflow developed in this work is introduced in Subsection \ref{workflow_overview}.
Subsection~\ref{mtp_overview} contains information on the MTP construction. The concept of extrapolation grade is described in Subsection~\ref{extrapolation_grade}, and the active learning scheme is presented in Subsection~\ref{sec:al_overview}.
Then, in Section~\ref{results}, we validate our methodology on ground-state structure prediction for CO/Pd(111), NO/Pd(100), NH$_3$/Cu(100), C$_6$H$_6$/Ag(111), and CH$_2$CO/Rh(211) systems.
Finally, in Section~\ref{discussions}, we discuss limitations and possible extensions of this work. Section~\ref{conclusion_and_outline} summarizes our findings.

\section{Methods}\label{Method}

\subsection{Workflow overview}\label{workflow_overview}

\begin{figure}
    \centering
    \includegraphics[width=0.5\textwidth]{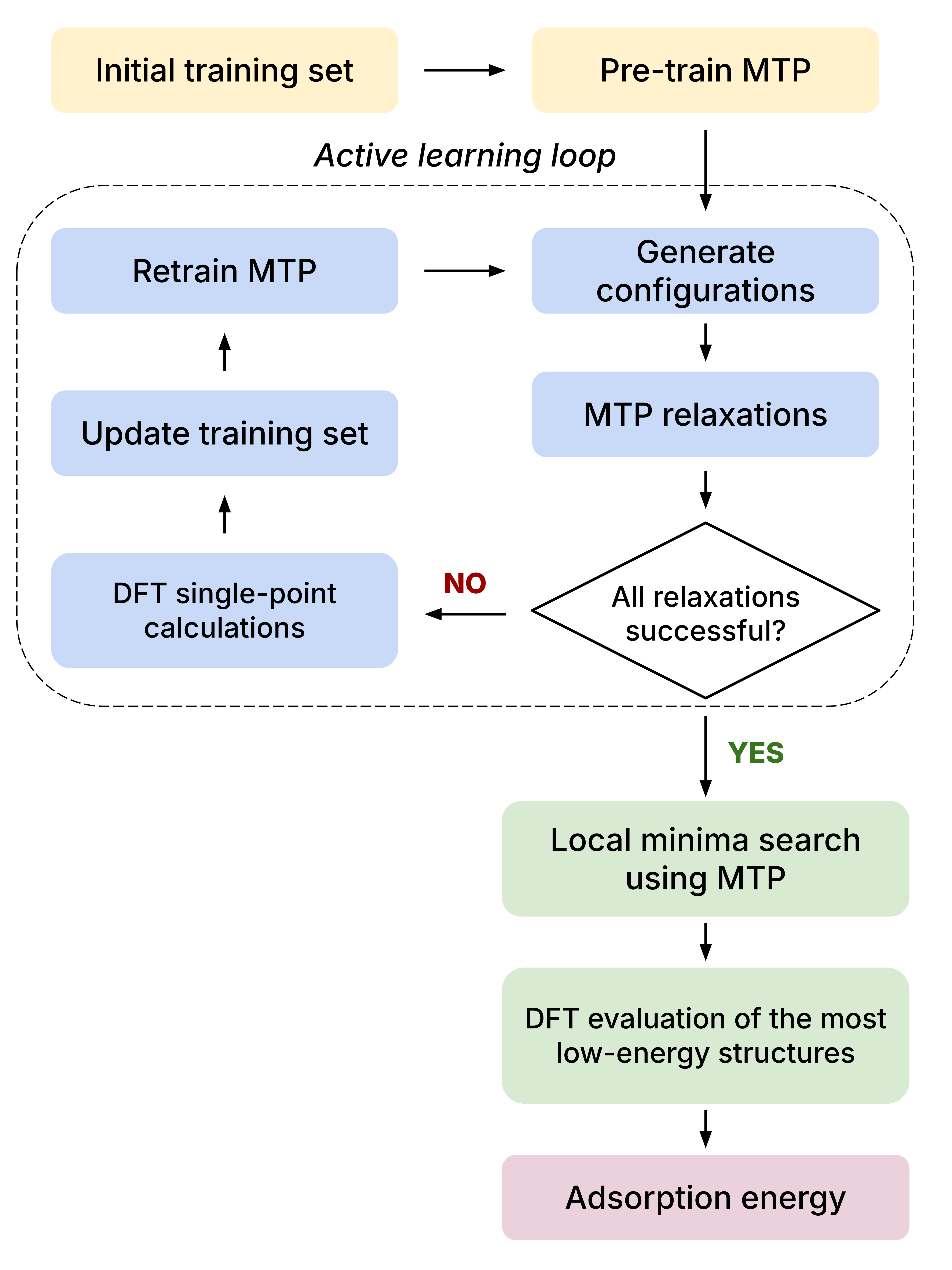}
    \caption{General scheme of the global optimization of surface adsorbate geometries. Initially, MTP is pre-trained on \textit{ab initio} molecular dynamics data or a set of single-point calculations. Then, molecules are randomly adsorbed on the surface and relaxed using MTP, with the slab fixed. If not all relaxations are successful, single-point calculations are provided for unknown structures, and the training set is updated. Subsequently, we retrain MTP and repeat the aforementioned steps until all relaxations are successfully completed. Thereafter, we randomly generate thousands of structures and relax them using MTP, which has been trained using the most recent data. Among these structures, we select the most energetically favorable structures and evaluate them using DFT. Finally, we identify the structure with the lowest energy, which we refer to as the ground state structure, and calculate its adsorption energy.}
    \label{fig:workflow}
\end{figure}

In this work, we follow the workflow shown in Fig. \ref{fig:workflow}.
The workflow starts with the pre-training of MTP, for which the training set is collected through one of the following ways.
First, the training set can be obtained through \textit{ab initio} molecular dynamics, in which the slab remains fixed while the molecule moves.
A less expensive approach to creating an initial training set involves various initialized structures with molecules on the surface and performing single-point calculations within the DFT framework.
Both methods are equally suitable in our case. 

After pre-training MTP, the active learning algorithm is initiated. This algorithm is described in detail in Subsection \ref{sec:al_overview}. During active learning, we perform structural optimization within the LAMMPS package \cite{LAMMPS}, while keeping the cell volume and the slab fixed.
In this step, MTP is used as a model for interatomic interactions. 
Here, we monitor that, at each step of relaxation, the potential does not extrapolate beyond the training set.
If the potential demonstrates significant extrapolation on the predictions of energies, it cannot be used for structural optimization purposes.
Therefore, once the potential approaches a risky extrapolation region, 
the relaxation process is halted, and the structures for which the potential extrapolates are added to the training set.
Next, DFT calculations are conducted for the structures that have been added to the training set, and the potential is re-fitted. This process, which involves relaxation with extrapolation control, selection of extrapolative structures, DFT calculations for them, and re-fitting the potential, is repeated until no extrapolative structures appear during the relaxation. 

When active learning has ended, the MTP is obtained, thereby enabling the calculation of energies and forces of structures with any possible molecule orientation on the surface.
Therefore, we can then perform thousands of MTP-based relaxations of structures with randomly initialized molecule on the surface.
Subsequently, dozens of structures with the lowest energy are selected and single-point DFT calculations are performed to evaluate the energies of the systems. The final step involves selecting the most promising structure with the lowest energy, as determined by DFT evaluation, as the ground-state structure.

\subsection{Moment Tensor Potential}\label{mtp_overview}

In this work, we used Moment Tensor Potentials (MTPs) implemented in the MLIP-2 package~\cite{Novikov2021} to perform structure optimization. In the scope of MTP, the potential energy of an atomic system is defined as a sum of the energies of atomic environments of the individual atoms:

\[
    E^{\text{MTP}} = \sum_{i=1}^{N} V({\mathfrak{\bm n}}_{i}),
\]
where the index $i$ labels $N$ atoms of the system, ${\mathfrak{\bm n}}_{i}$ denotes the local atomic neighborhood around atom \textit{i} within a certain cut-off radius $R_\text{cut}$, and the function $V$ is the energy of atomic neighborhood: 
\[
    V({\mathfrak{\bm n}}_{i}) = \sum_{\alpha} \xi_{\alpha} B_{\alpha}({\mathfrak{\bm n}}_{i}).
\]
Here, $\xi_{\alpha}$ are the linear parameters to be fitted and $B_{\alpha}({\mathfrak{\bm n}}_{i})$ are the basis functions that will be defined below. As fundamental symmetry requirements, all descriptors for atomic environment have to be invariant to translation, rotation, and permutation with respect to the atomic indexing. Moment tensors descriptors $M_{\mu, \nu}$ satisfy these requirements and are used as representations of atomic environments: 
\[
    M_{\mu, \nu}\left({\mathfrak{\bm n}}_{i}\right)=\sum_j f_\mu\left(\left|{\bm r}_{i j}\right|, z_i, z_j\right) \underbrace{{\bm r}_{i j} \otimes \ldots \otimes \bm{r}_{i j}}_{\nu \text { times }},
\]
where the index $j$ goes through all the neighbors of atom $i$. The symbol ``$\otimes$'' stands for the outer product of vectors, thus ${\bm r}_{i j} \otimes \cdots \otimes {\bm r}_{i j}$ is the tensor of rank $\nu$ encoding the angular part. 
The function $f_\mu$ represents the radial component of the moment tensor:
\[
f_\mu\left(\left|{\bm r}_{i j}\right|, z_i, z_j\right)=\sum_{\beta} c_{\mu, z_i, z_j}^{(\beta)} Q^{(\beta)}\left(\left|{\bm r}_{i j}\right|\right),
\]
where $z_i$ and $z_j$ denote the atomic species of atoms $i$ and $j$, respectively, ${\bm r}_{ij}$ describes the positioning of atom $j$ relative to atom $i$, $c_{\mu, z_i, z_j}^{(\beta)}$ are the radial parameters to be fitted and 
\[
Q^{(\beta)}\left(\left|{\bm r}_{i j}\right|\right)=T^{(\beta)}\left(\left|{\bm r}_{i j}\right|\right)\left(R_{\text {cut }}-\left|{\bm r}_{i j}\right|\right)^2
\]
are the radial functions consisting of the Chebyshev polynomials $T^{(\beta)}\left(\left|{\bm r}_{i j}\right|\right)$ on the interval $[R_\text{min},  R_\text{cut}]$ and the term $\left(R_\text{cut} - \left|{\bm r}_{i j}\right|\right)^2$ that is introduced to ensure a smooth cut-off to zero. We emphasize that the number of the radial parameters scales quadratically with the number of atomic species in structures. The descriptors $M_{\mu, \nu}$ taking $\nu$ equal to $0, 1, 2, \ldots$ are tensors of different ranks that allow one to define basis functions as all possible contractions of these tensors to a scalar, for instance:
\[
\begin{aligned}
& B_0\left({\mathfrak{\bm n}}_{i}\right)=M_{0,0}\left({\mathfrak{\bm n}}_{i}\right), \\
& B_1\left({\mathfrak{\bm n}}_{i}\right)=M_{0,1}\left({\mathfrak{\bm n}}_{i}\right) \cdot M_{0,1}\left({\mathfrak{\bm n}}_{i}\right), \\
& B_2\left({\mathfrak{\bm n}}_{i}\right)=\left(M_{2,2}\left({\mathfrak{\bm n}}_{i}\right) M_{2,1}\left({\mathfrak{\bm n}}_i\right)\right) \cdot  M_{0,1}\left({\mathfrak{\bm n}}_{i}\right), \\
& \ldots
\end{aligned}
\]

\noindent However, the number of contractions yielding a scalar, i.e., basis functions $B_{\alpha}$ is infinite. In order to restrict the number of basis functions in the MTP functional form, we introduce the level of moment tensor descriptors ${\rm lev}M_{\mu, \nu}$ = 2 + 4$\mu$ + $\nu$. If $B_{\alpha}$ is obtained from $M_{\mu_1, \nu_1}$, $M_{\mu_2, \nu_2}$ ,
$\dots$, then ${\rm lev}B_{\alpha}$ = $(2 + 4\mu_1 + \nu_1)$ + $(2 + 4\mu_2 + \nu_2)$ + $\dots$. By including all the basis functions with ${\rm lev}B_{\alpha} \leq d$, we obtain the MTP of level $d$ including a finite number of the basis functions $B_{\alpha}$. We denote the total set of parameters to be found by ${\bm \theta} = (\{ \xi_{\alpha} \}, \{ c^{(\beta)}_{\mu, z_i, z_j} \})$ and the MTP energy of a structure $E^{\text{MTP}} = E({\bm \theta})$.

\subsection{Extrapolation grade of structures}\label{extrapolation_grade}

For automatically constructing a training set for MTP, we calculate the so-called extrapolation grade of structures obtained during the structural relaxation.
For estimating the extrapolation grade of structures we conduct the following steps. Assume we have $K$ structures in an initial training set with energies $E^{\rm DFT}_k$, forces ${\bm f}^{\rm DFT}_{i,k}$, and stresses $\sigma^{\rm DFT}_{i,k}$, $k = 1, \ldots, K$ calculated within the DFT framework. We start with fitting an initial MTP, i.e., with finding the optimal parameters ${\bm{\bar{\theta}}}$ by solving the following minimization problem:
\begin{equation} \notag
\begin{array}{c}
\displaystyle
\sum \limits_{k=1}^K \Bigl[ w_{\rm e} \left(E^{\rm DFT}_k - E_k({\bm {\theta}}) \right)^2 + w_{\rm f} \sum_{i=1}^N \left| {\bm f}^{\rm DFT}_{i,k} - {\bm f}_{i,k}({\bm {\theta}}) \right|^2 
\\
\displaystyle
+ w_{\rm s} \sum_{i=1}^6 \left| \sigma^{\rm DFT}_{i,k} - \sigma_{i,k}({\bm {\theta}}) \right|^2 \Bigr] \to \operatorname{min},
\end{array}
\end{equation}
where $w_{\rm e}$, $w_{\rm f}$, and $w_{\rm s}$ are non-negative weights expressing the importance of energies, forces, and stresses in the above minimization problem. 

After finding the optimal parameters ${\bm{\bar{\theta}}}$ we compose the following matrix
\[
\mathsf{B}=\left(\begin{matrix}
\frac{\partial E_1\left( {\bm{\bar{\theta}}} \right)}{\partial \theta_1} & \ldots & \frac{\partial E_1\left( {\bm{\bar{\theta}}} \right)}{\partial \theta_m} \\
\vdots & \ddots & \vdots \\
\frac{\partial E_K\left( {\bm{\bar{\theta}}} \right)}{\partial \theta_1} & \ldots & \frac{\partial E_K\left( {\bm{\bar{\theta}}} \right)}{\partial \theta_m} \\
\end{matrix}\right),
\]
where each row corresponds to a particular structure. Next, we construct a subset of structures yielding the most linearly independent rows (physically it means geometrically different structures) in $\mathsf{B}$. This is equivalent to finding a square $m \times m$ submatrix $\mathsf{A}$ of the matrix $\mathsf{B}$ of maximum volume (maximal value of $|{\rm det(\mathsf{A})}|$). To achieve this, we use the so-called maxvol algorithm \cite{zamarashkin2010-maxvol}. The $m$ structures in the matrix $\mathsf{A}$ is called an active set. To determine whether a given structure $\bm x^*$ obtained during structural relaxation is representative, we calculate the extrapolation grade $\gamma(\bm x^*)$ defined as
\begin{equation} \label{Grade}
\begin{array}{c}
\displaystyle
\gamma(\bm x^*) = \max_{1 \leq j \leq m} (|c_j|), ~\rm{where}
\\
\displaystyle
{\bm c} = \left( \dfrac{\partial E}{\partial \theta_1} (\bm{\bar{\theta}}, \bm x^*) \ldots \dfrac{\partial E}{\partial \theta_m} (\bm{\bar{\theta}}, \bm x^*) \right) \mathsf{A}^{-1}.
\end{array}
\end{equation}
This grade defines the maximal factor by which the determinant $|{\rm det(\mathsf{A})}|$ can increase if ${\bm x^*}$ is added to the training set. Thus, if the structure $\bm x^*$ is a candidate for adding to the training set then $\gamma(\bm x^*) \geq \gamma_{\rm th}$, where $\gamma_{\rm th} \geq 1$ is an adjustable threshold parameter which controls the value of permissible extrapolation. Otherwise, the structure is not representative. In this work we used $\gamma_{\rm th} = 2.1$ and once the extrapolation grade exceeds 10 the relaxation was stopped. Such values were chosen based on the previous benchmarks~\cite{Podryabinkin2017,Novikov2021}.

\subsection{Active learning algorithm}\label{sec:al_overview}

Next, we describe the active learning scheme for an automatic construction of a training set and fit MTP for predicting ground-state adsorbate-surface structures.

\textbf{Input to the algorithm:}
\begin{enumerate}
\item A set of candidate structures, among which we expect to find
the ground-state structures. We assume that we do not know the preferred orientation of the molecule on the surface, that is why we need our potential to be able to predict the energy of structures of all possible orientations of the molecule at all possible sites.    
\item Pre-trained MTP.

\item A quantum mechanical model to provide DFT calculations. In this work, we used DFT as implemented in Vienna \textit{ab initio} simulation package (VASP) \cite{vasp}.

\item Two thresholds $\gamma_{tsh} = 2$ and $\Gamma_{tsh} = 10$, where $\Gamma_{tsh} > \gamma_{tsh} > 1$. If during active learning the extrapolation grade $\gamma$ becomes greater than 1, the algorithm makes the decision: if $\gamma > \gamma_{tsh}$, but $\gamma < \Gamma_{tsh}$, then the configuration is added to the preselected set for further DFT calculations. If $\gamma > \Gamma_{tsh}$, then relaxation is terminated, because in that case we cannot make reliable predictions of energy, forces, and stresses for such configuration.

\end{enumerate}

\textbf{Step 1:} For each candidate structure, we perform the structure relaxation with the current MTP. There can be two outcomes of the relaxation: first, the relaxation completed successfully and we get the equilibrium structure as the result, and, second, the relaxation is terminated as MTP needs to extrapolate. More detailed information on two scenarios is as follows:

\begin{itemize}
\item The structure successfully converges to an equilibrium configuration and on each configuration from the relaxation trajectory, the MTP does not significantly extrapolate, i.e., the extrapolation grade of each intermediate configuration is less than $\Gamma_{tsh}$.

\item At some step of relaxation we obtain a configuration with the extrapolation grade exceeding $\Gamma_{tsh}$. This means that MTP cannot provide a reliable prediction as it extrapolates significantly on this configuration and needs to be retrained with more \textit{ab initio}  data. We then terminate the relaxation. The last and all previous configurations with the grade exceeding $\gamma_{tsh}$ are added to the \textit{preselected set}.
\end{itemize}

\textbf{Step 2:} We select a possibly smaller number of configurations from the preselected set that will be added to the training set. We use the active learning algorithm to select the most representative configurations, according to the D-optimality criterion \cite{GUBAEV2019148}. After that we conduct \textit{ab initio} calculations of energies, forces and stresses for selected configurations and add them to the training set.

\textbf{Step 3:} Fit MTP on the updated training set. As the size of the training set grows on each iteration of the algorithm, this step will take more and more time during each iteration, but still this time is much smaller than \textit{ab initio} calculations.

\textbf{Step 4:} Repeat steps 1-3 unless all relaxations have successfully converged to the respective equilibrium configurations.

\subsection{Computational details}

Surface slabs were constructed from (3$\times$3) surface supercells with a thickness of four metal layers. We calculate adsorption energies ($E_{\text{ads}}$) as

\begin{equation*}
    E_{\text{ads}} = E_{\text{slab+mol}} - E_{\text{slab}} - E_{\text{mol}},
\end{equation*}
where $E_{\text{slab+mol}}$ is the energy of the combined surface and adsorbate system, $E_{\text{slab}}$ is the energy of the clean slab, and $E_{\text{mol}}$ is energy of the molecule in a gas phase.

The MTP utilized in this study is trained using data computed within the DFT framework.
All DFT calculations were carried out using Vienna \textit{ab initio} simulations package (VASP)~\cite{vasp} with the projector-augmented wave method~\cite{Kresse1999}.
The Perdew-Burke-Ernzerhof generalized gradient approximation (PBE-GGA)~\cite{Perdew1996} was employed for the exchange–correlation functional.
The Grimme D3 dispersion correction~\cite{Grimme2010} was applied to take into account van der Waals interactions between surfaces and adsorbed molecules.
Dipole correction was included in the direction of the surface normal.
For single-point calculations, we used a plane-wave energy cut-off of 500~eV.
Convergence with respect to the k-points density using total energy as a marker in the two-dimensional Brillouin zone has been checked. The convergence parameters for energy and forces were set to $1\times 10^{-5}$~eV and $1\times 10^{-4}$~eV/\AA \, respectively.

In this work, we used MTP of the 12-th level, i.e., with 248 parameters for systems with 3 atom types and 417 parameters for systems with 4 atoms types. We took a cut-off radius of 5 \AA ~in order to ensure a non-zero interaction between molecules and surfaces. All structures were generated using ASE \cite{ase-paper} and ACAT \cite{acat}, visualized via OVITO software \cite{ovito}.

\section{Results}\label{results}

\begin{figure*}
 \centering
\subfigure{\includegraphics[width=0.32\textwidth]{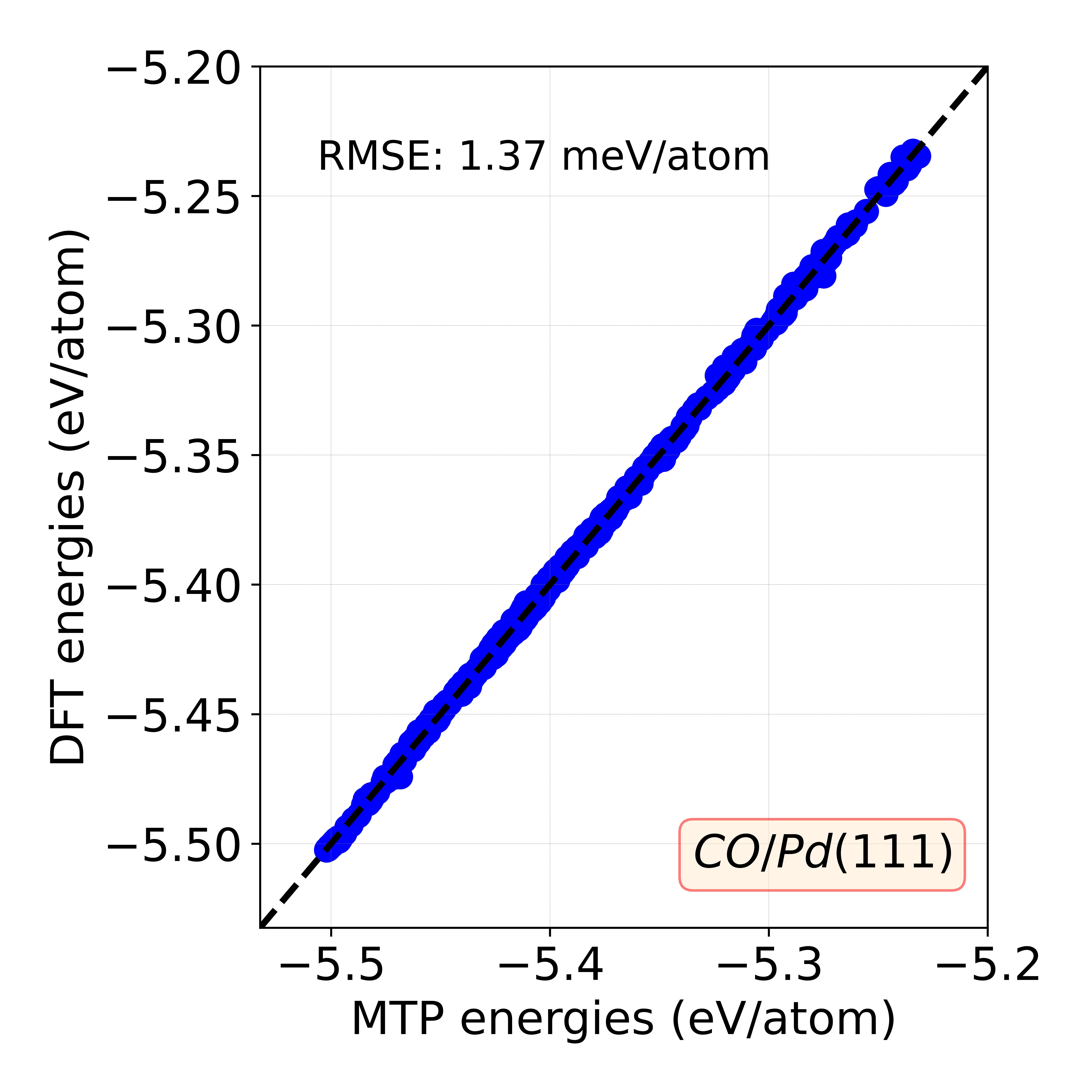}}
\subfigure{\includegraphics[width=0.32\textwidth]{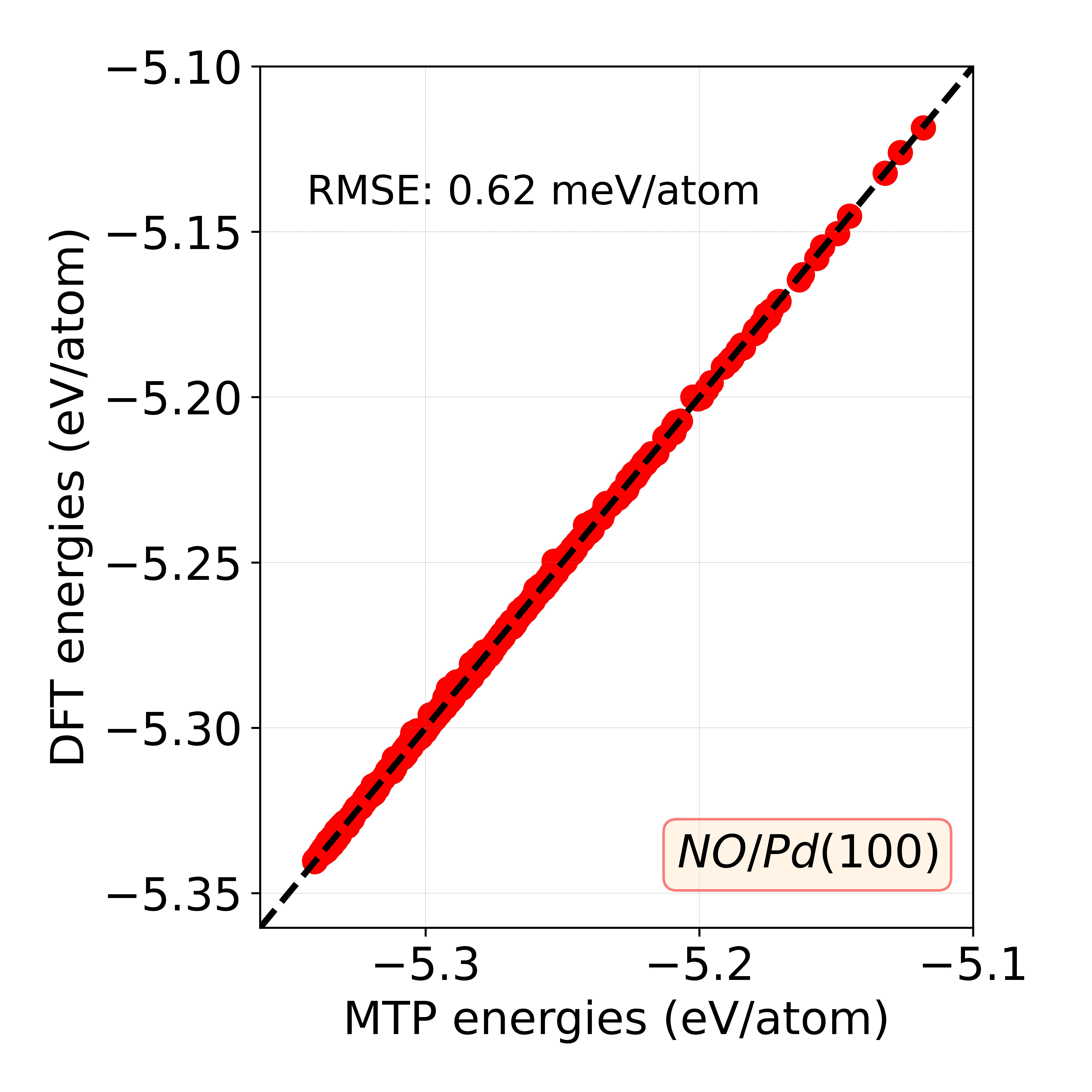}}
\subfigure{\includegraphics[width=0.32\textwidth]{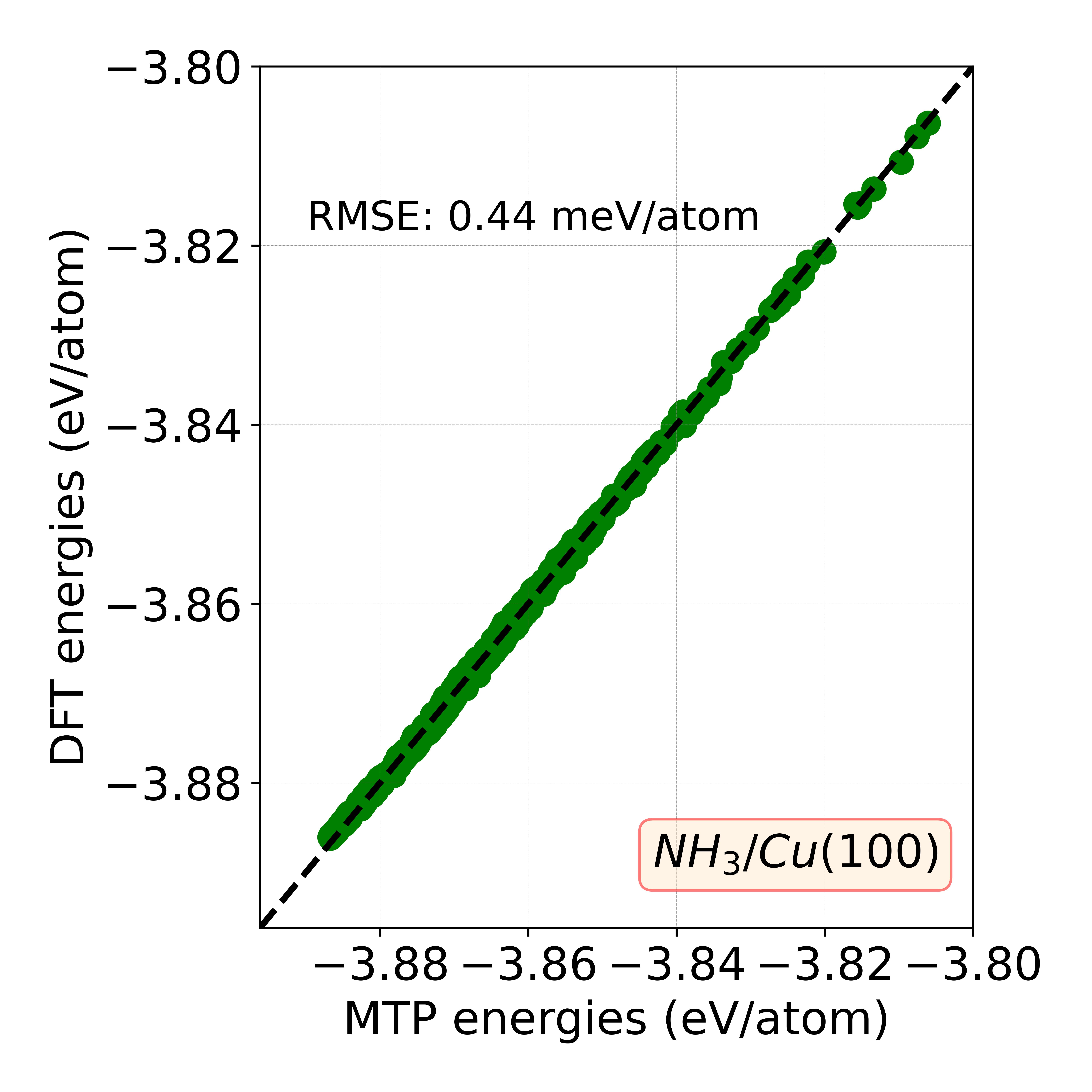}}
\subfigure{\includegraphics[width=0.32\textwidth]{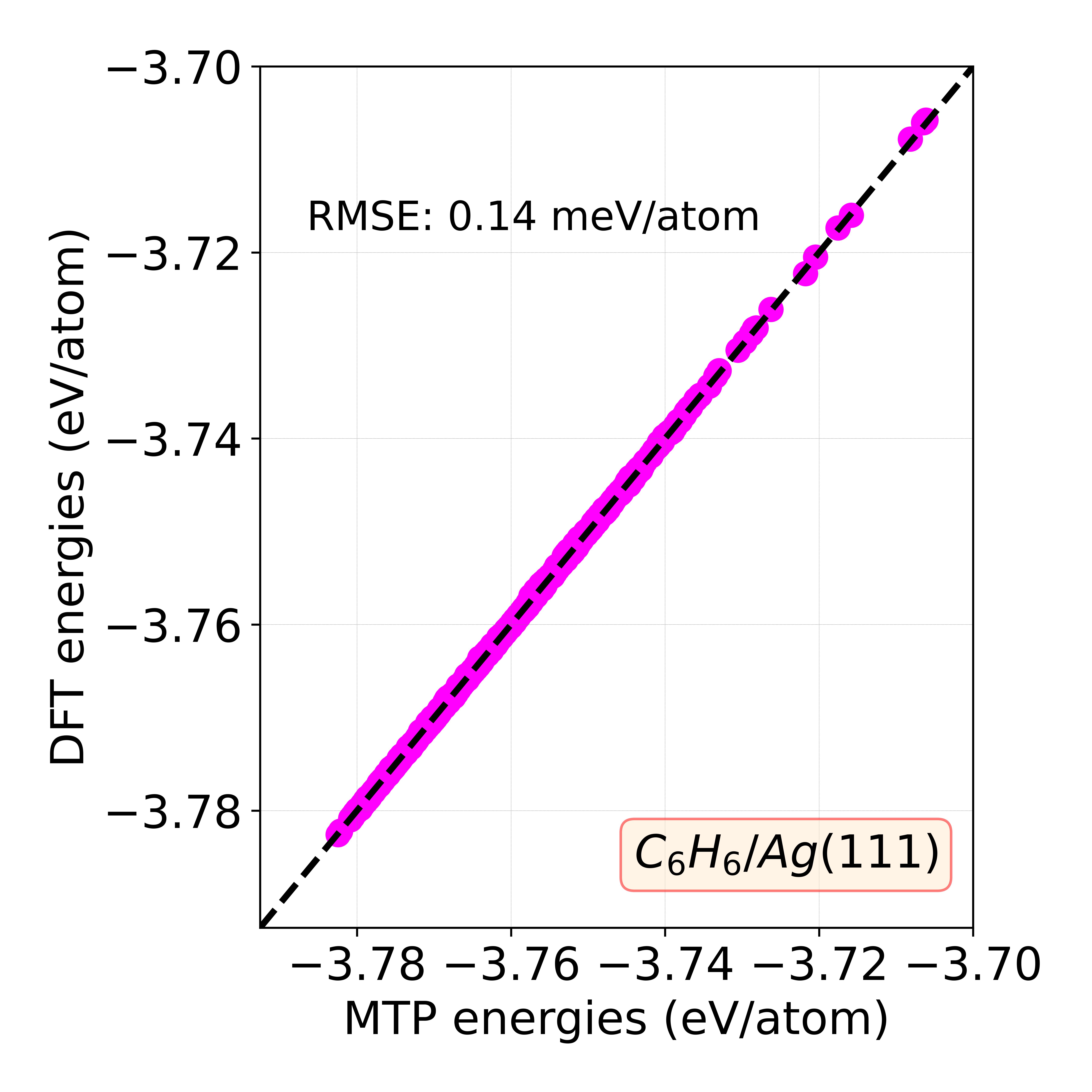}}
\subfigure{\includegraphics[width=0.32\textwidth]{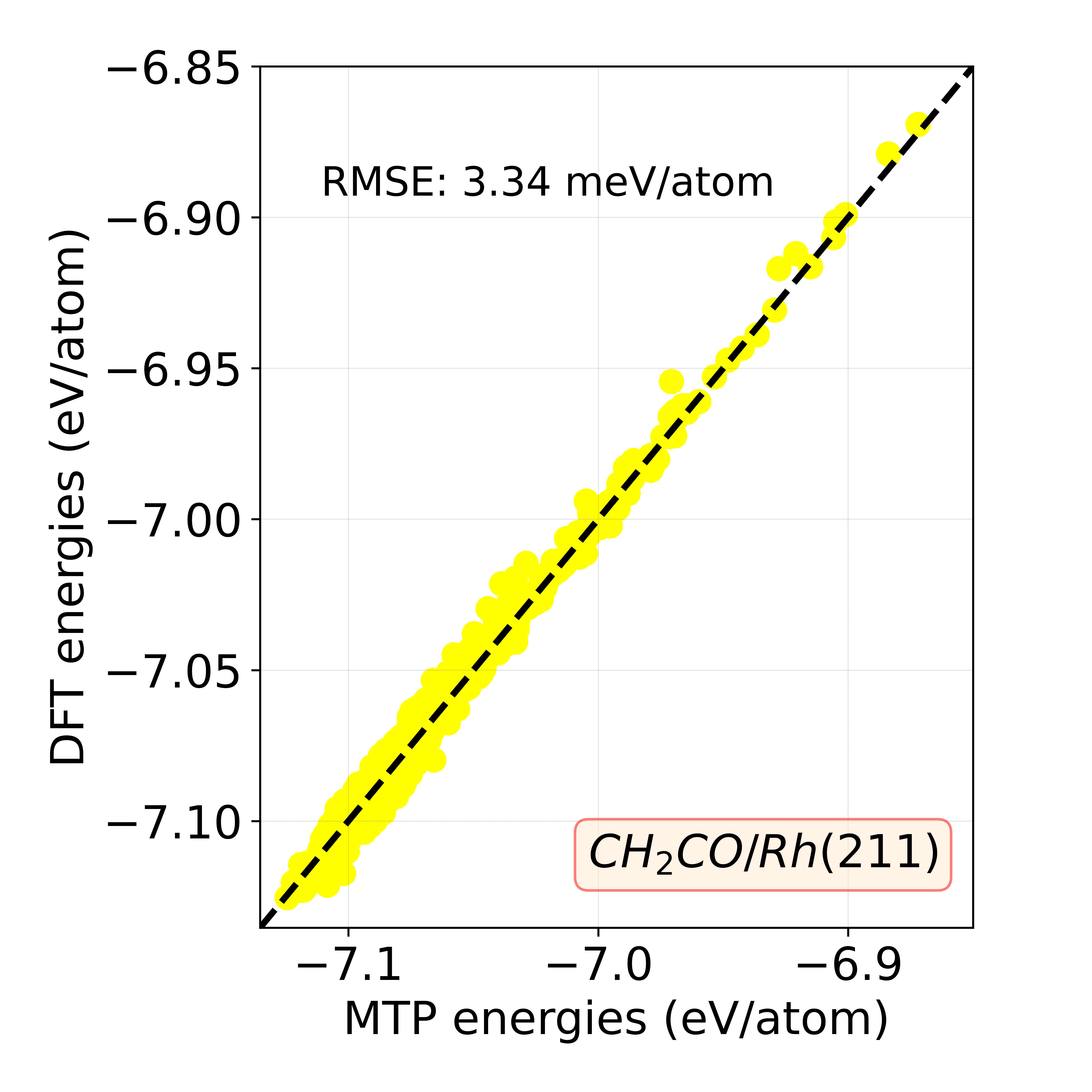}}
  \caption{Comparison between DFT-calculated and MTP-calculated energies of the structures in training datasets. Number of structures in the training datasets are specified in teh Table \ref{tab:test_train_errors}.}
  \label{fig:mlip_dft}
 \end{figure*}

To illustrate the applicability of the proposed methodology, we considered a set of systems featuring small to mid-sized adsorbates on metallic surfaces with different orientations: CO/Pd(111), NO/Pd(100), NH$_3$/Cu(100), C$_6$H$_6$/Ag(111), and CH$_2$CO/Rh(211). The potential fitting and active learning procedures were consistent across all cases. The information regarding the potential complexity, training sets, and errors on energies, forces, and stresses is presented in Table~\ref{tab:test_train_errors}.
Furthermore, in Fig. \ref{fig:mlip_dft}, we compare the MTP and DFT energies of the structures in training sets for all systems.
We demonstrate that the energies derived from MTP for the structures within the training set match those obtained through DFT, with root mean square errors no greater than 3.34 meV/atom.

\begin{table*}[ht]
 \caption{Number of parameters in MTP used for the five systems, number of samples, and mean absolute errors on energies, forces and stresses.} 
\centering
\begin{tabular*}{\textwidth}{@{\extracolsep{\fill}}cccccc}
\hline
& CO/Pd(111) & NO/Pd(100) & NH$_3$/Cu(100) & C$_6$H$_6$/Ag(111) & CH$_2$CO/Rh(211) \\ \hline
\begin{tabular}[c]{@{}c@{}}\# Parameters\\ in MTP\end{tabular}                  & 248     & 248     & 248 & 248  & 417 \\ \hline
\begin{tabular}[c]{@{}c@{}}\# Samples \\ in the \\ training set\end{tabular}    & 799  & 1226 & 824 & 797 & 1399 \\ \hline
\begin{tabular}[c]{@{}c@{}}Train MAE \\ on energy \\ (meV/atom)\end{tabular}    & 1.08    & 0.46 & 0.36 & 0.10 & 2.32 \\ \hline
\begin{tabular}[c]{@{}c@{}}Train MAE \\ on forces\\ (meV/\AA)\end{tabular} & 22.9    & 10.8 &  7.2 & 5.9 & 31.9 \\ \hline
\begin{tabular}[c]{@{}c@{}}Train MAE \\ on stresses\\ (GPa)\end{tabular} & 0.022    & 0.014 & 0.019 & 0.076 & 0.026 \\ 
\hline
\end{tabular*}
\label{tab:test_train_errors}
\end{table*}

\subsection{First example: CO/Pd(111)}

In this section, we present the results of applying our method for calculating the adsorption energy of the CO molecule on the Pd(111) surface. This system is simple enough for the initial validation of our computations, as it involves a linear molecule as the adsorbate. As discussed earlier, each calculation is initiated without prior assumptions regarding the molecule's preferred orientation on the surface. Although the literature indicates that CO orients towards Pd(111) with carbon closer to the surface \cite{CO_orienton_Pd111, araujo2022adsorption}, the MTP has to learn all possible orientations. 

\begin{figure}
    \centering
    \includegraphics[width=0.5\textwidth]{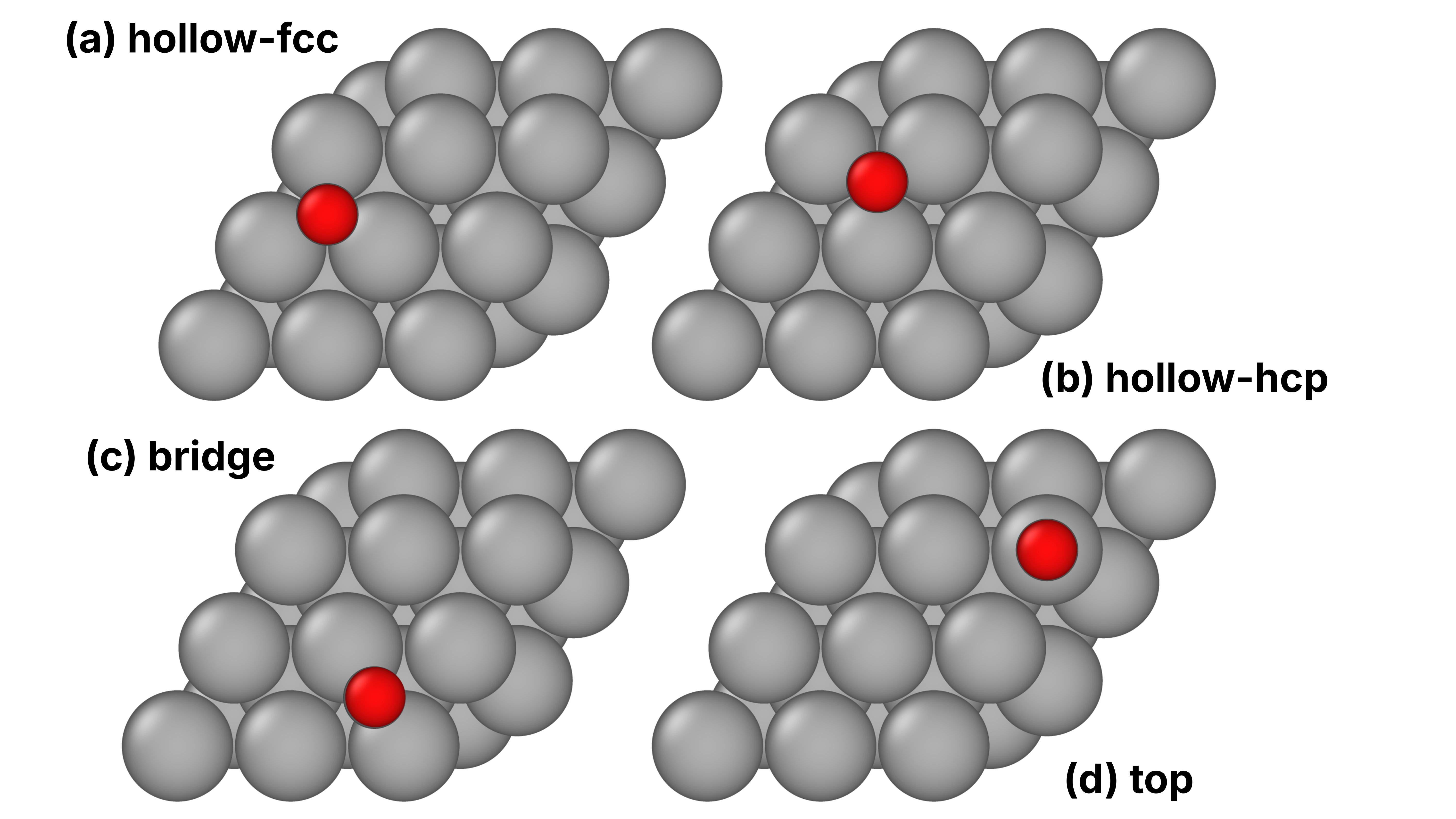}
    \caption{Visualization of different adsorption sites of the CO molecule on the Pd(111) surface -- hollow fcc and hcp, bridge and top. The CO molecule is oriented perpendicular to the slab, with carbon closer to the surface.}
    \label{fig:ads_cites}
\end{figure}

\begin{figure}
    \centering
    \includegraphics[width=0.5\textwidth]{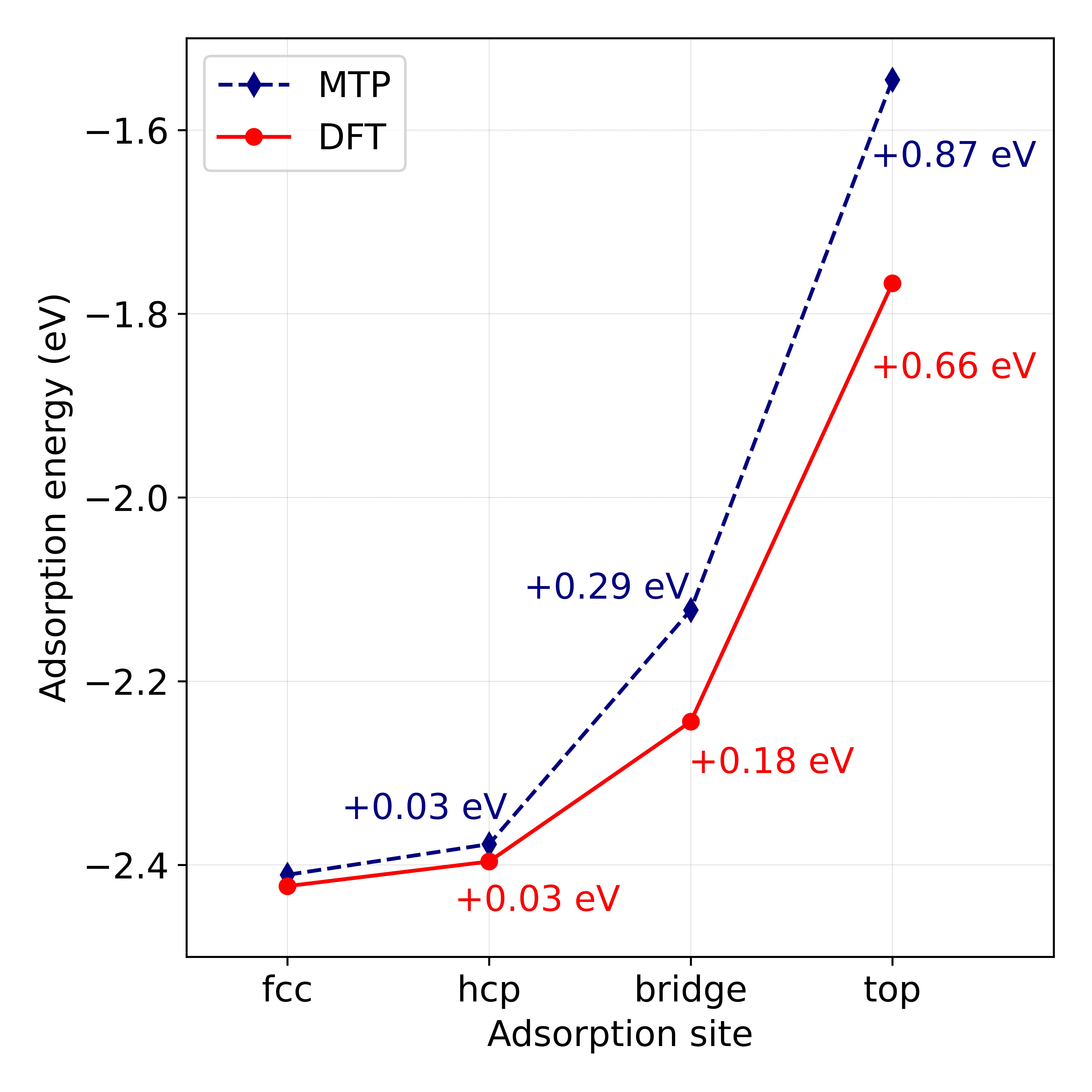}
    \caption{Calculated adsorption energies for the CO molecule on the Pd(111) surface are presented as a function of different adsorption sites. All energies are referenced to the most stable configuration, which is the hollow-fcc site. The DFT results in our work are in agreement with the work \cite{CO_Pd111_fcc_hcp}, and MTP reproduces them qualitatively well.}
    \label{fig:ads_CO_Pd111}
\end{figure}

Following our workflow, shown in Fig. \ref{fig:workflow}, we obtained the adsorption energy of $-$2.410~eV using MTP and $-$2.427~eV after DFT evaluation.
These findings indicate an agreement between the values obtained with MTP and the \textit{ab initio} approach, as well as with the literature, which reports an adsorption energy of $-$2.220 $\pm$ 0.603~eV \cite{araujo2022adsorption} for the same adsorption site.
As a result, CO preferentially adsorbs at the hollow-fcc site.
However, the theoretical results from the literature~\cite{CO_Pd111_fcc_hcp} indicate a minor difference in formation energy between hollow-fcc and hollow-hcp sites, specifically $0.03$ eV.
Thus, we checked whether our MTP could differentiate among various adsorption sites: hollow-fcc, hollow-hcp, bridge, and top, as illustrated in Fig. \ref{fig:ads_cites}.
In Fig. \ref{fig:ads_CO_Pd111}, we observe that our MTP can distinguish different adsorption sites even with a small energy difference of 0.03~eV between hollow-fcc and hollow-hcp.
This is not a coincidence.
As demonstrated in Fig. \ref{fig:mlip_dft}, the root mean square error between the energies calculated using the MTP and those obtained from DFT for the structures in the CO/Pd(111) dataset is 1.37 meV/atom.
The DFT results for these adsorption energies in Fig. \ref{fig:ads_CO_Pd111} align with theoretical predictions from the literature \cite{CO_Pd111_fcc_hcp}, which include hollow-fcc as the ground state, with energy differences of $+$0.03~eV for hollow-hcp, $+$0.20~eV for bridge, and $+$0.65~eV for top sites. It is notable that the greater the deviation of a structure's energy from the global minimum, the larger the difference between the results obtained using MTP and DFT. It happens because during relaxation, molecules try to reach the global energy minimum; thus, configurations within the training set that describe the potential energy surface near this minimum are more common. 

\subsection{Other results}

In addition to the CO/Pd(111) system, we validated our approach on NO/Pd(100), NH$_3$/Cu(100), C$_6$H$_6$/Ag(111), and CH$_2$CO/Rh(211).

\begin{figure}
    \centering
    \includegraphics[width=0.5\textwidth]{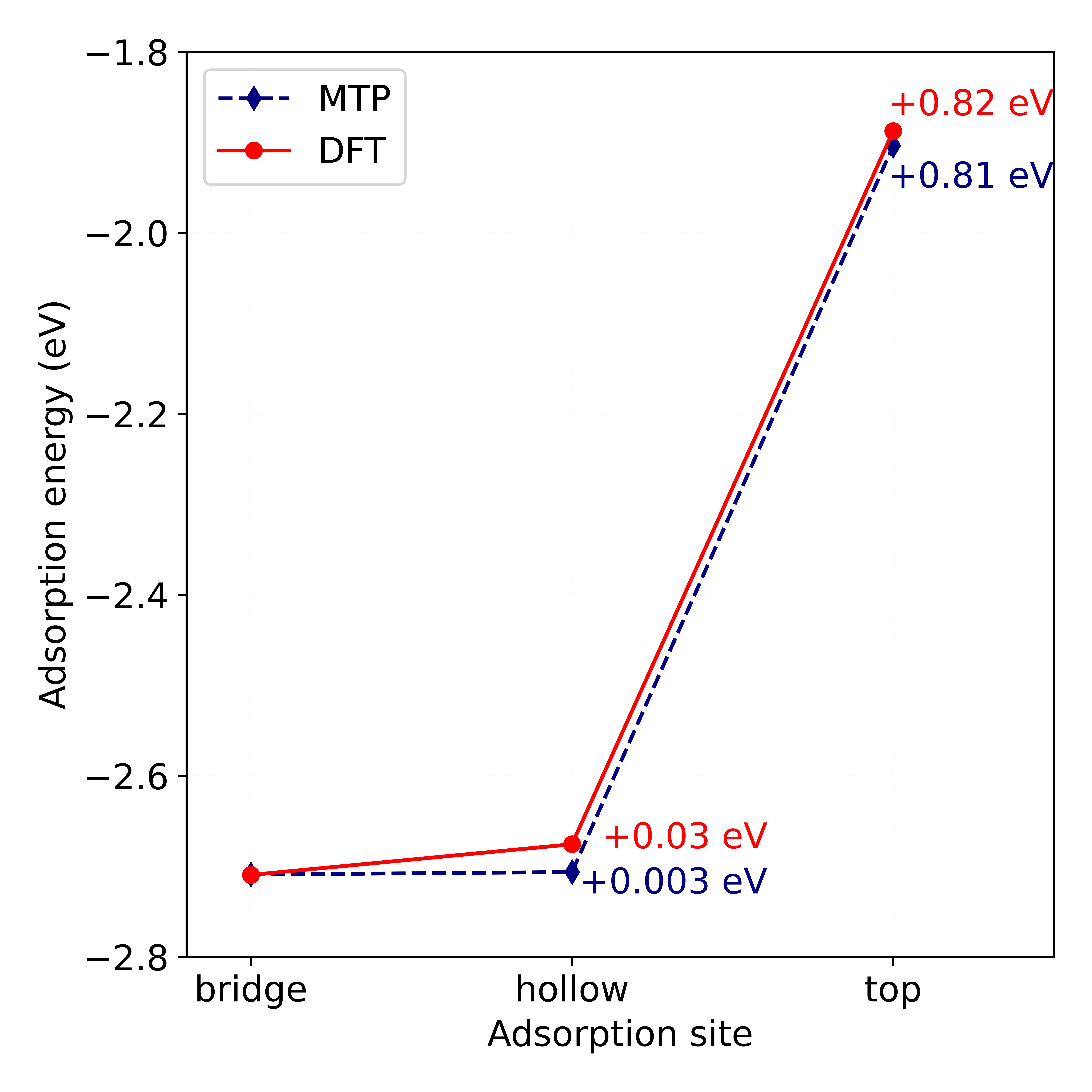}
    \caption{Calculated adsorption energies for the NO molecule on the Pd(100) surface are presented as a function of different adsorption sites. All energies are referenced to the most stable configuration, which is the bridge site. We see that MTP reproduces DFT results qualitatively and quantitatively well.}
    \label{fig:ads_NO_Pd100}
\end{figure}

The bridge adsorption site was identified as the most favorable for NO/Pd(100) system by our algorithm.
However, the literature indicates that NO prefers to adsorb on both the hollow and bridge sites on Pd(100) \cite{no_1, no_2, no_3}.
Therefore, to verify whether our algorithm correctly identified the position of the adsorbate, we computed the adsorption energies for all sites within this system, as shown in Fig. \ref{fig:ads_NO_Pd100}.
According to our DFT calculations, the adsorption energy for the bridge site is $-$2.710~eV, which is by 0.03~eV lower than that for the hollow site. As we can see, our algorithm can distinguish this difference.

\begin{figure}
    \centering
    \includegraphics[width=0.5\textwidth]{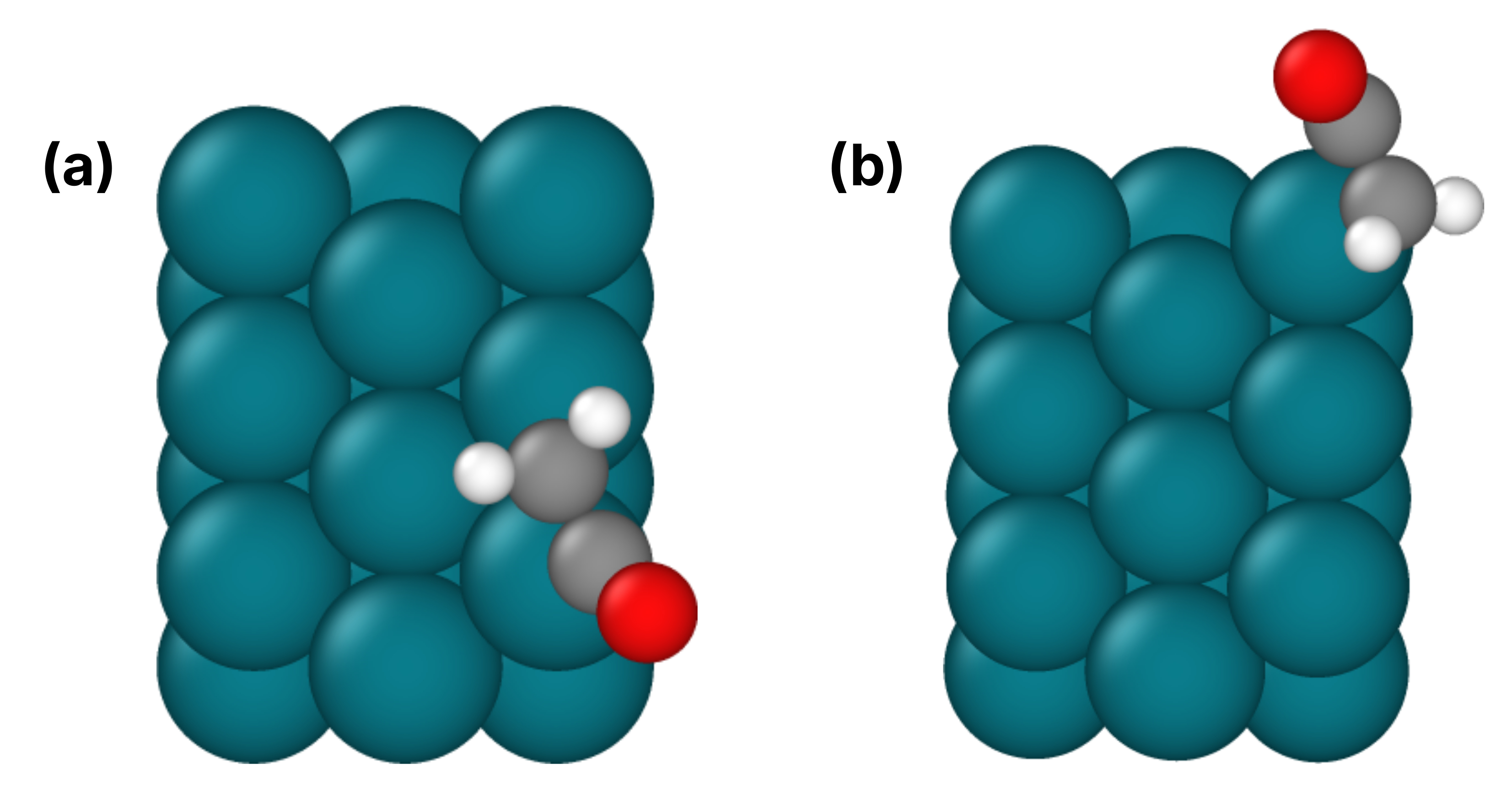}
    \caption{ (a) The ground-state structure of CH$_2$CO/Rh(211) system in our work. (b) The ground-state structure of CH$_2$CO/Rh(211) system in the work \cite{GAP_ads}. The (b) structure was taken from the dataset \cite{gap_dataset}. In both cases, the CH$_2$CO adsorbed on the step edges.}
    \label{fig:gap_compare}
\end{figure}

Then we also considered the adsorption of NH$_3$ on Cu(100), C$_6$H$_6$ on Ag(111), and CH$_2$CO on Rh(211). Our findings indicate that NH$_3$ prefers to adsorb at the top site, while C$_6$H$_6$ adsorb on hcp-30 site.
The identification of a precise adsorption site for CH$_2$CO/Rh(211) proved challenging due to its location along the step edge in our work.
To validate our findings, we compared our optimized structure with those reported in the work \cite{GAP_ads,gap_dataset} as a global minimum configuration.
As illustrated in Fig. \ref{fig:gap_compare}, both the structures in our work and those from \cite{GAP_ads} prefer the adsorption along the step edges. It is important to note that we took the ground-state structure from \cite{GAP_ads}, which was obtained using the plane-wave Quantum Espresso code \cite{QS} with the BEEF-vdW functional \cite{PhysRevB.85.235149_BElib}. Therefore, we performed single-point DFT calculations for this structure and its slab using our settings with the PBE+D3 functional.
All the calculated adsorption energies in our work are compiled in Table \ref{tab:ads_en}.

\begin{table*}
  \caption{Adsorption energies obtained by MTP and evaluated by PBE+D3 approach, reference energies from literature also calculated by PBE+D3 approach, and adsorption sites.}
  \label{tab:ads_en}
  \begin{tabular*}{\textwidth}{@{\extracolsep{\fill}}cccccc}
    \hline
    & CO/Pd(111) & NO/Pd(100) & NH$_3$/Cu(100) & C$_6$H$_6$/Ag(111) & CH$_2$CO/Rh(211) \\ 
    \hline
\begin{tabular}[c]{@{}c@{}}\ Adsorption energy\\ MTP (eV)\end{tabular}
    & $-$2.410 & $-$2.709 & $-$0.864 & $-$0.720 & $-$1.922 \\  \hline
\begin{tabular}[c]{@{}c@{}}\ Adsorption energy\\ DFT (eV)\end{tabular}
    & $-$2.427 & $-$2.710 & $-$0.893 & $-$0.726 & $-$1.972 \\ \hline
\begin{tabular}[c]{@{}c@{}}\ Adsorption energy\\ DFT in lit. (eV)\end{tabular}
    & $-$2.220 $\pm$ 0.603 \cite{araujo2022adsorption} & $-$2.428 $\pm$ 0.603$^{\ast}$ \cite{araujo2022adsorption} & $-$0.681 $\pm$ 0.247 \cite{araujo2022adsorption} & $-$0.86 \cite{C6H6_Ag111_hcp30} & $-$2.102$^{\ast \ast}$ \cite{GAP_ads}  \\ \hline
\begin{tabular}[c]{@{}c@{}}\ Site \end{tabular}
    & fcc & bridge & top & hcp-30 & step edge \\ 
    \hline
  \end{tabular*}
$^{\ast}$We compare the value for the bridge site in our work with the value for the hollow site from the work \cite{araujo2022adsorption}, because according to the results in Fig. \ref{fig:ads_NO_Pd100} both values are close to each other and can be compared within the error obtained in the work \cite{araujo2022adsorption} for the PBE+D3 approach. \\
$^{\ast \ast}$We took the structure from the dataset \cite{gap_dataset} and provided single-point calculations for the whole structure, clean slab and molecule with our DFT settings.  
\end{table*}

\section{Discussion}\label{discussions}

Our results show a great applicability of the proposed methodology for accelerating global optimization of surface adsorbate geometries using machine learning interatomic potentials. Combined with the active learning, our approach enables to decrease the computational cost of data collection and eliminates the necessity for manual data processing. As a result, our findings agree well with the existing literature concerning the preferred adsorption sites of diverse molecules and the adsorption energies of the considered systems. 

In this study, we did not concentrate on hyperparameter optimization for our model.
However, in order to reduce training errors, it would be beneficial to use MTPs of higher levels, such as 16, 20 and 24, which would also improve the accuracy of adsorption energy calculations.
Furthermore, there are several limitations that should be discussed.
While the results obtained for all structures are promising and suggest the potential application of MTP for global optimization of molecular configurations on surfaces, it is important to note that these systems are relatively simple.
Future work will involve applying our methodology to more complex systems, such as adsorption on alloy nanoparticles.
Suggested workflow can be also used to speed up global optimization of surface adsorbate geometries even on larger systems. From our perspective, one of the major challenges ahead is associated with the limitation of DFT calculations as the system size increases.
Therefore, we would like to continue working on the development of our model using the MLIP-3 software \cite{MLIP-3}, which might allow one to provide active learning on the atomic neighborhoods of a possibly large atomistic simulation.

Last but not least, all the calculations described here were performed at zero temperature and pressure.
In future research, we intend to consider systems under catalytic conditions varying temperature and pressure, particularly focusing on catalyst reconstruction in electrocatalysis, as the reconstructed structure has the capacity to promote or hinder electrochemical performance.
It would be useful to study how to achieve the desired active surface that promotes highly catalytic activity through reconstruction \cite{surf_rec}.

\section{Conclusions}\label{conclusion_and_outline}

We have developed an algorithm for accelerating global optimization of the position of the surface adsorbate molecule, using the moment tensor potential as a model for interatomic interactions. 
The active-learning methodology, a key element of our algorithm, facilitates the generation of accurate potentials with minimal data utilization.
We demonstrated that MTP can achieve both speed and accuracy in optimizing the adsorbate-surface configurations, obtaining results that are comparable to those obtained by DFT calculations.
The methodology developed in our work will be adapted for future investigations of more complex systems, such as adsorption on alloy nanoparticles and amorphous solids.

\section{Author contributions}

O.K. performed calculations and active training of MTP. N.R. initiated the study and supervised the research. A.S. supervised the research and provided the funding. All authors participated in the manuscript writing, reviewing, and editing.

\section{Conflicts of interest}
We have no conflicts of interest to disclose.

\section{Acknowledgements}

Authors acknowledge funding from the Russian Science Foundation (Project No. 23-13-00332).

\printcredits

\bibliographystyle{model1-num-names}

\bibliography{main}

\begin{thebibliography}{42}
\expandafter\ifx\csname natexlab\endcsname\relax\def\natexlab#1{#1}\fi
\providecommand{\url}[1]{\texttt{#1}}
\providecommand{\href}[2]{#2}
\providecommand{\path}[1]{#1}
\providecommand{\DOIprefix}{doi:}
\providecommand{\ArXivprefix}{arXiv:}
\providecommand{\URLprefix}{URL: }
\providecommand{\Pubmedprefix}{pmid:}
\providecommand{\doi}[1]{\href{http://dx.doi.org/#1}{\path{#1}}}
\providecommand{\Pubmed}[1]{\href{pmid:#1}{\path{#1}}}
\providecommand{\bibinfo}[2]{#2}
\ifx\xfnm\relax \def\xfnm[#1]{\unskip,\space#1}\fi
\bibitem[{Busacca et~al.(2011)Busacca, Fandrick, Song, and Senanayake}]{pharma}
\bibinfo{author}{C.~A. Busacca}, \bibinfo{author}{D.~R. Fandrick}, \bibinfo{author}{J.~J. Song}, \bibinfo{author}{C.~H. Senanayake},
\newblock \bibinfo{title}{The growing impact of catalysis in the pharmaceutical industry},
\newblock \bibinfo{journal}{Advanced Synthesis \& Catalysis} \bibinfo{volume}{353} (\bibinfo{year}{2011}) \bibinfo{pages}{1825--1864}.
\bibitem[{Liu et~al.(2018)Liu, Ma, Li, Wang, Xiao, and Li}]{ammonia}
\bibinfo{author}{J.-C. Liu}, \bibinfo{author}{X.-L. Ma}, \bibinfo{author}{Y.~Li}, \bibinfo{author}{Y.-G. Wang}, \bibinfo{author}{H.~Xiao}, \bibinfo{author}{J.~Li},
\newblock \bibinfo{title}{Heterogeneous fe3 single-cluster catalyst for ammonia synthesis via an associative mechanism},
\newblock \bibinfo{journal}{Nature communications} \bibinfo{volume}{9} (\bibinfo{year}{2018}) \bibinfo{pages}{1610}.
\bibitem[{Cheng et~al.(2017)Cheng, Kang, King, Subramanian, Zhou, Zhang, and Wang}]{syngas}
\bibinfo{author}{K.~Cheng}, \bibinfo{author}{J.~Kang}, \bibinfo{author}{D.~L. King}, \bibinfo{author}{V.~Subramanian}, \bibinfo{author}{C.~Zhou}, \bibinfo{author}{Q.~Zhang}, \bibinfo{author}{Y.~Wang},
\newblock \bibinfo{title}{Chapter three - advances in catalysis for syngas conversion to hydrocarbons},
\newblock volume~\bibinfo{volume}{60} of \textit{\bibinfo{series}{Advances in Catalysis}}, \bibinfo{publisher}{Academic Press}, \bibinfo{year}{2017}, pp. \bibinfo{pages}{125--208}.
\bibitem[{N{\o}rskov et~al.(2009)N{\o}rskov, Bligaard, Rossmeisl, and Christensen}]{cat_1}
\bibinfo{author}{J.~K. N{\o}rskov}, \bibinfo{author}{T.~Bligaard}, \bibinfo{author}{J.~Rossmeisl}, \bibinfo{author}{C.~H. Christensen},
\newblock \bibinfo{title}{Towards the computational design of solid catalysts},
\newblock \bibinfo{journal}{Nature chemistry} \bibinfo{volume}{1} (\bibinfo{year}{2009}) \bibinfo{pages}{37--46}.
\bibitem[{Bruix et~al.(2019)Bruix, Margraf, Andersen, and Reuter}]{cat_2}
\bibinfo{author}{A.~Bruix}, \bibinfo{author}{J.~T. Margraf}, \bibinfo{author}{M.~Andersen}, \bibinfo{author}{K.~Reuter},
\newblock \bibinfo{title}{First-principles-based multiscale modelling of heterogeneous catalysis},
\newblock \bibinfo{journal}{Nature Catalysis} \bibinfo{volume}{2} (\bibinfo{year}{2019}) \bibinfo{pages}{659--670}.
\bibitem[{Su et~al.(2020)Su, Zhang, Wang, Liu, Muravev, Alexopoulos, Filot, Vlachos, and Hensen}]{cat_3}
\bibinfo{author}{Y.-Q. Su}, \bibinfo{author}{L.~Zhang}, \bibinfo{author}{Y.~Wang}, \bibinfo{author}{J.-X. Liu}, \bibinfo{author}{V.~Muravev}, \bibinfo{author}{K.~Alexopoulos}, \bibinfo{author}{I.~A. Filot}, \bibinfo{author}{D.~G. Vlachos}, \bibinfo{author}{E.~J. Hensen},
\newblock \bibinfo{title}{Stability of heterogeneous single-atom catalysts: a scaling law mapping thermodynamics to kinetics},
\newblock \bibinfo{journal}{npj Computational Materials} \bibinfo{volume}{6} (\bibinfo{year}{2020}) \bibinfo{pages}{144}.
\bibitem[{Zitnick et~al.(2020)Zitnick, Chanussot, Das, Goyal, Heras-Domingo, Ho, Hu, Lavril, Palizhati, Riviere, Shuaibi, Sriram, Tran, Wood, Yoon, Parikh, and Ulissi}]{open_catalyst}
\bibinfo{author}{C.~L. Zitnick}, \bibinfo{author}{L.~Chanussot}, \bibinfo{author}{A.~Das}, \bibinfo{author}{S.~Goyal}, \bibinfo{author}{J.~Heras-Domingo}, \bibinfo{author}{C.~Ho}, \bibinfo{author}{W.~Hu}, \bibinfo{author}{T.~Lavril}, \bibinfo{author}{A.~Palizhati}, \bibinfo{author}{M.~Riviere}, \bibinfo{author}{M.~Shuaibi}, \bibinfo{author}{A.~Sriram}, \bibinfo{author}{K.~Tran}, \bibinfo{author}{B.~Wood}, \bibinfo{author}{J.~Yoon}, \bibinfo{author}{D.~Parikh}, \bibinfo{author}{Z.~Ulissi}, \bibinfo{title}{An introduction to electrocatalyst design using machine learning for renewable energy storage}, \bibinfo{year}{2020}. \URLprefix \url{https://arxiv.org/abs/2010.09435}. \href{http://arxiv.org/abs/2010.09435}{\tt arXiv:2010.09435}.
\bibitem[{Chanussot et~al.(2021)Chanussot, Das, Goyal, Lavril, Shuaibi, Riviere, Tran, Heras-Domingo, Ho, Hu, Palizhati, Sriram, Wood, Yoon, Parikh, Zitnick, and Ulissi}]{OC20}
\bibinfo{author}{L.~Chanussot}, \bibinfo{author}{A.~Das}, \bibinfo{author}{S.~Goyal}, \bibinfo{author}{T.~Lavril}, \bibinfo{author}{M.~Shuaibi}, \bibinfo{author}{M.~Riviere}, \bibinfo{author}{K.~Tran}, \bibinfo{author}{J.~Heras-Domingo}, \bibinfo{author}{C.~Ho}, \bibinfo{author}{W.~Hu}, \bibinfo{author}{A.~Palizhati}, \bibinfo{author}{A.~Sriram}, \bibinfo{author}{B.~Wood}, \bibinfo{author}{J.~Yoon}, \bibinfo{author}{D.~Parikh}, \bibinfo{author}{C.~L. Zitnick}, \bibinfo{author}{Z.~Ulissi},
\newblock \bibinfo{title}{Open catalyst 2020 (oc20) dataset and community challenges},
\newblock \bibinfo{journal}{ACS Catalysis} \bibinfo{volume}{11} (\bibinfo{year}{2021}) \bibinfo{pages}{6059--6072}.
\bibitem[{Tran et~al.(2023)Tran, Lan, Shuaibi, Wood, Goyal, Das, Heras-Domingo, Kolluru, Rizvi, Shoghi, Sriram, Therrien, Abed, Voznyy, Sargent, Ulissi, and Zitnick}]{OC22}
\bibinfo{author}{R.~Tran}, \bibinfo{author}{J.~Lan}, \bibinfo{author}{M.~Shuaibi}, \bibinfo{author}{B.~M. Wood}, \bibinfo{author}{S.~Goyal}, \bibinfo{author}{A.~Das}, \bibinfo{author}{J.~Heras-Domingo}, \bibinfo{author}{A.~Kolluru}, \bibinfo{author}{A.~Rizvi}, \bibinfo{author}{N.~Shoghi}, \bibinfo{author}{A.~Sriram}, \bibinfo{author}{F.~Therrien}, \bibinfo{author}{J.~Abed}, \bibinfo{author}{O.~Voznyy}, \bibinfo{author}{E.~H. Sargent}, \bibinfo{author}{Z.~Ulissi}, \bibinfo{author}{C.~L. Zitnick},
\newblock \bibinfo{title}{The open catalyst 2022 (oc22) dataset and challenges for oxide electrocatalysts},
\newblock \bibinfo{journal}{ACS Catalysis} \bibinfo{volume}{13} (\bibinfo{year}{2023}) \bibinfo{pages}{3066--3084}.
\bibitem[{Ringe(2023)}]{descriptor_1}
\bibinfo{author}{S.~Ringe},
\newblock \bibinfo{title}{The importance of a charge transfer descriptor for screening potential co2 reduction electrocatalysts},
\newblock \bibinfo{journal}{Nature Communications} \bibinfo{volume}{14} (\bibinfo{year}{2023}) \bibinfo{pages}{2598}.
\bibitem[{Huang et~al.(2021)Huang, Li, Zhao, Chen, Bu, and Cheng}]{descriptor_2}
\bibinfo{author}{H.-C. Huang}, \bibinfo{author}{J.~Li}, \bibinfo{author}{Y.~Zhao}, \bibinfo{author}{J.~Chen}, \bibinfo{author}{Y.-X. Bu}, \bibinfo{author}{S.-B. Cheng},
\newblock \bibinfo{title}{Adsorption energy as a promising single-parameter descriptor for single atom catalysis in the oxygen evolution reaction},
\newblock \bibinfo{journal}{J. Mater. Chem. A} \bibinfo{volume}{9} (\bibinfo{year}{2021}) \bibinfo{pages}{6442--6450}.
\bibitem[{Sutton and Vlachos(2012)}]{descriptor_3}
\bibinfo{author}{J.~E. Sutton}, \bibinfo{author}{D.~G. Vlachos},
\newblock \bibinfo{title}{A theoretical and computational analysis of linear free energy relations for the estimation of activation energies},
\newblock \bibinfo{journal}{ACS Catalysis} \bibinfo{volume}{2} (\bibinfo{year}{2012}) \bibinfo{pages}{1624--1634}.
\bibitem[{Pablo-Garc{\'\i}a et~al.(2023)Pablo-Garc{\'\i}a, Morandi, Vargas-Hern{\'a}ndez, Jorner, Ivkovi{\'c}, L{\'o}pez, and Aspuru-Guzik}]{GNN}
\bibinfo{author}{S.~Pablo-Garc{\'\i}a}, \bibinfo{author}{S.~Morandi}, \bibinfo{author}{R.~A. Vargas-Hern{\'a}ndez}, \bibinfo{author}{K.~Jorner}, \bibinfo{author}{{\v{Z}}.~Ivkovi{\'c}}, \bibinfo{author}{N.~L{\'o}pez}, \bibinfo{author}{A.~Aspuru-Guzik},
\newblock \bibinfo{title}{Fast evaluation of the adsorption energy of organic molecules on metals via graph neural networks},
\newblock \bibinfo{journal}{Nature Computational Science} \bibinfo{volume}{3} (\bibinfo{year}{2023}) \bibinfo{pages}{433--442}.
\bibitem[{Lan et~al.(2023)Lan, Palizhati, Shuaibi, Wood, Wander, Das, Uyttendaele, Zitnick, and Ulissi}]{AdsorbML}
\bibinfo{author}{J.~Lan}, \bibinfo{author}{A.~Palizhati}, \bibinfo{author}{M.~Shuaibi}, \bibinfo{author}{B.~Wood}, \bibinfo{author}{B.~Wander}, \bibinfo{author}{A.~Das}, \bibinfo{author}{M.~Uyttendaele}, \bibinfo{author}{C.~Zitnick}, \bibinfo{author}{Z.~Ulissi},
\newblock \bibinfo{title}{Adsorbml: a leap in efficiency for adsorption energy calculations using generalizable machine learning potentials},
\newblock \bibinfo{journal}{npj Computational Materials} \bibinfo{volume}{9} (\bibinfo{year}{2023}).
\bibitem[{Jung et~al.(2023)Jung, Sauerland, Stocker, Reuter, and Margraf}]{GAP_ads}
\bibinfo{author}{H.~Jung}, \bibinfo{author}{L.~Sauerland}, \bibinfo{author}{S.~Stocker}, \bibinfo{author}{K.~Reuter}, \bibinfo{author}{J.~T. Margraf},
\newblock \bibinfo{title}{Machine-learning driven global optimization of surface adsorbate geometries},
\newblock \bibinfo{journal}{npj Computational Materials} \bibinfo{volume}{9} (\bibinfo{year}{2023}) \bibinfo{pages}{114}.
\bibitem[{Shapeev(2016)}]{Shapeev2016}
\bibinfo{author}{A.~V. Shapeev},
\newblock \bibinfo{title}{Moment tensor potentials: A class of systematically improvable interatomic potentials},
\newblock \bibinfo{journal}{Multiscale Modeling \& Simulation} \bibinfo{volume}{14} (\bibinfo{year}{2016}) \bibinfo{pages}{1153--1173}.
\bibitem[{Gubaev et~al.(2019)Gubaev, Podryabinkin, Hart, and Shapeev}]{GUBAEV2019148}
\bibinfo{author}{K.~Gubaev}, \bibinfo{author}{E.~V. Podryabinkin}, \bibinfo{author}{G.~L. Hart}, \bibinfo{author}{A.~V. Shapeev},
\newblock \bibinfo{title}{Accelerating high-throughput searches for new alloys with active learning of interatomic potentials},
\newblock \bibinfo{journal}{Computational Materials Science} \bibinfo{volume}{156} (\bibinfo{year}{2019}) \bibinfo{pages}{148--156}.
\bibitem[{Rybin et~al.(2024)Rybin, Novikov, and Shapeev}]{mol_crystals}
\bibinfo{author}{N.~Rybin}, \bibinfo{author}{I.~S. Novikov}, \bibinfo{author}{A.~Shapeev},
\newblock \bibinfo{title}{Accelerating structure prediction of molecular crystals using actively trained moment tensor potential},
\newblock \bibinfo{journal}{arXiv preprint arXiv:2410.03484}  (\bibinfo{year}{2024}).
\bibitem[{Podryabinkin et~al.(2019)Podryabinkin, Tikhonov, Shapeev, and Oganov}]{boron}
\bibinfo{author}{E.~V. Podryabinkin}, \bibinfo{author}{E.~V. Tikhonov}, \bibinfo{author}{A.~V. Shapeev}, \bibinfo{author}{A.~R. Oganov},
\newblock \bibinfo{title}{Accelerating crystal structure prediction by machine-learning interatomic potentials with active learning},
\newblock \bibinfo{journal}{Phys. Rev. B} \bibinfo{volume}{99} (\bibinfo{year}{2019}) \bibinfo{pages}{064114}.
\bibitem[{Podryabinkin and Shapeev(2017)}]{Podryabinkin2017}
\bibinfo{author}{E.~V. Podryabinkin}, \bibinfo{author}{A.~V. Shapeev},
\newblock \bibinfo{title}{Active learning of linearly parametrized interatomic potentials},
\newblock \bibinfo{journal}{Computational Materials Science} \bibinfo{volume}{140} (\bibinfo{year}{2017}) \bibinfo{pages}{171--180}.
\bibitem[{Plimpton(1995)}]{LAMMPS}
\bibinfo{author}{S.~Plimpton},
\newblock \bibinfo{title}{Fast parallel algorithms for short-range molecular dynamics},
\newblock \bibinfo{journal}{Journal of Computational Physics} \bibinfo{volume}{117} (\bibinfo{year}{1995}) \bibinfo{pages}{1--19}.
\bibitem[{Novikov et~al.(2020)Novikov, Gubaev, Podryabinkin, and Shapeev}]{Novikov2021}
\bibinfo{author}{I.~S. Novikov}, \bibinfo{author}{K.~Gubaev}, \bibinfo{author}{E.~V. Podryabinkin}, \bibinfo{author}{A.~V. Shapeev},
\newblock \bibinfo{title}{The mlip package: moment tensor potentials with mpi and active learning},
\newblock \bibinfo{journal}{Machine Learning: Science and Technology} \bibinfo{volume}{2} (\bibinfo{year}{2020}) \bibinfo{pages}{025002}.
\bibitem[{Goreinov et~al.(2010)Goreinov, Oseledets, Savostyanov, Tyrtyshnikov, and Zamarashkin}]{zamarashkin2010-maxvol}
\bibinfo{author}{S.~A. Goreinov}, \bibinfo{author}{I.~V. Oseledets}, \bibinfo{author}{D.~V. Savostyanov}, \bibinfo{author}{E.~E. Tyrtyshnikov}, \bibinfo{author}{N.~L. Zamarashkin},
\newblock \bibinfo{title}{How to find a good submatrix},
\newblock in: \bibinfo{booktitle}{Matrix Methods: Theory, Algorithms And Applications: Dedicated to the Memory of Gene Golub}, \bibinfo{publisher}{World Scientific}, \bibinfo{year}{2010}, pp. \bibinfo{pages}{247--256}.
\bibitem[{Kresse and Furthm\"uller(1996)}]{vasp}
\bibinfo{author}{G.~Kresse}, \bibinfo{author}{J.~Furthm\"uller},
\newblock \bibinfo{title}{Efficient iterative schemes for ab initio total-energy calculations using a plane-wave basis set},
\newblock \bibinfo{journal}{Phys. Rev. B} \bibinfo{volume}{54} (\bibinfo{year}{1996}) \bibinfo{pages}{11169--11186}.
\bibitem[{Kresse and Joubert(1999)}]{Kresse1999}
\bibinfo{author}{G.~Kresse}, \bibinfo{author}{D.~Joubert},
\newblock \bibinfo{title}{From ultrasoft pseudopotentials to the projector augmented-wave method},
\newblock \bibinfo{journal}{Phys. Rev. B} \bibinfo{volume}{59} (\bibinfo{year}{1999}) \bibinfo{pages}{1758--1775}.
\bibitem[{Perdew et~al.(1996)Perdew, Burke, and Ernzerhof}]{Perdew1996}
\bibinfo{author}{J.~P. Perdew}, \bibinfo{author}{K.~Burke}, \bibinfo{author}{M.~Ernzerhof},
\newblock \bibinfo{title}{Generalized gradient approximation made simple},
\newblock \bibinfo{journal}{Phys. Rev. Lett.} \bibinfo{volume}{77} (\bibinfo{year}{1996}) \bibinfo{pages}{3865--3868}.
\bibitem[{Grimme et~al.(2010)Grimme, Antony, Ehrlich, and Krieg}]{Grimme2010}
\bibinfo{author}{S.~Grimme}, \bibinfo{author}{J.~Antony}, \bibinfo{author}{S.~Ehrlich}, \bibinfo{author}{H.~Krieg},
\newblock \bibinfo{title}{{A consistent and accurate ab initio parametrization of density functional dispersion correction (DFT-D) for the 94 elements H-Pu}},
\newblock \bibinfo{journal}{The Journal of Chemical Physics} \bibinfo{volume}{132} (\bibinfo{year}{2010}) \bibinfo{pages}{154104}.
\bibitem[{Larsen et~al.(2017)Larsen, Mortensen, Blomqvist, Castelli, Christensen, Dułak, Friis, Groves, Hammer, Hargus, Hermes, Jennings, Jensen, Kermode, Kitchin, Kolsbjerg, Kubal, Kaasbjerg, Lysgaard, Maronsson, Maxson, Olsen, Pastewka, Peterson, Rostgaard, Schiøtz, Schütt, Strange, Thygesen, Vegge, Vilhelmsen, Walter, Zeng, and Jacobsen}]{ase-paper}
\bibinfo{author}{A.~H. Larsen}, \bibinfo{author}{J.~J. Mortensen}, \bibinfo{author}{J.~Blomqvist}, \bibinfo{author}{I.~E. Castelli}, \bibinfo{author}{R.~Christensen}, \bibinfo{author}{M.~Dułak}, \bibinfo{author}{J.~Friis}, \bibinfo{author}{M.~N. Groves}, \bibinfo{author}{B.~Hammer}, \bibinfo{author}{C.~Hargus}, \bibinfo{author}{E.~D. Hermes}, \bibinfo{author}{P.~C. Jennings}, \bibinfo{author}{P.~B. Jensen}, \bibinfo{author}{J.~Kermode}, \bibinfo{author}{J.~R. Kitchin}, \bibinfo{author}{E.~L. Kolsbjerg}, \bibinfo{author}{J.~Kubal}, \bibinfo{author}{K.~Kaasbjerg}, \bibinfo{author}{S.~Lysgaard}, \bibinfo{author}{J.~B. Maronsson}, \bibinfo{author}{T.~Maxson}, \bibinfo{author}{T.~Olsen}, \bibinfo{author}{L.~Pastewka}, \bibinfo{author}{A.~Peterson}, \bibinfo{author}{C.~Rostgaard}, \bibinfo{author}{J.~Schiøtz}, \bibinfo{author}{O.~Schütt}, \bibinfo{author}{M.~Strange}, \bibinfo{author}{K.~S. Thygesen}, \bibinfo{author}{T.~Vegge}, \bibinfo{author}{L.~Vilhelmsen}, \bibinfo{author}{M.~Walter},
  \bibinfo{author}{Z.~Zeng}, \bibinfo{author}{K.~W. Jacobsen},
\newblock \bibinfo{title}{The atomic simulation environment—a python library for working with atoms},
\newblock \bibinfo{journal}{Journal of Physics: Condensed Matter} \bibinfo{volume}{29} (\bibinfo{year}{2017}) \bibinfo{pages}{273002}.
\bibitem[{Han et~al.(2023)Han, Lysgaard, Vegge, and Hansen}]{acat}
\bibinfo{author}{S.~Han}, \bibinfo{author}{S.~Lysgaard}, \bibinfo{author}{T.~Vegge}, \bibinfo{author}{H.~A. Hansen},
\newblock \bibinfo{title}{Rapid mapping of alloy surface phase diagrams via bayesian evolutionary multitasking},
\newblock \bibinfo{journal}{npj Computational Materials} \bibinfo{volume}{9} (\bibinfo{year}{2023}) \bibinfo{pages}{139}.
\bibitem[{Stukowski(2009)}]{ovito}
\bibinfo{author}{A.~Stukowski},
\newblock \bibinfo{title}{Visualization and analysis of atomistic simulation data with ovito–the open visualization tool},
\newblock \bibinfo{journal}{Modelling and Simulation in Materials Science and Engineering} \bibinfo{volume}{18} (\bibinfo{year}{2009}) \bibinfo{pages}{015012}.
\bibitem[{Hooshmand et~al.(2017)Hooshmand, Le, and Rahman}]{CO_orienton_Pd111}
\bibinfo{author}{Z.~Hooshmand}, \bibinfo{author}{D.~Le}, \bibinfo{author}{T.~S. Rahman},
\newblock \bibinfo{title}{Co adsorption on pd(111) at 0.5ml: A first principles study},
\newblock \bibinfo{journal}{Surface Science} \bibinfo{volume}{655} (\bibinfo{year}{2017}) \bibinfo{pages}{7--11}.
\bibitem[{Araujo et~al.(2022)Araujo, Rodrigues, Dos~Santos, and Pettersson}]{araujo2022adsorption}
\bibinfo{author}{R.~B. Araujo}, \bibinfo{author}{G.~L. Rodrigues}, \bibinfo{author}{E.~C. Dos~Santos}, \bibinfo{author}{L.~G. Pettersson},
\newblock \bibinfo{title}{Adsorption energies on transition metal surfaces: towards an accurate and balanced description},
\newblock \bibinfo{journal}{Nature Communications} \bibinfo{volume}{13} (\bibinfo{year}{2022}) \bibinfo{pages}{6853}.
\bibitem[{Sautet et~al.(2000)Sautet, Rose, Dunphy, Behler, and Salmeron}]{CO_Pd111_fcc_hcp}
\bibinfo{author}{P.~Sautet}, \bibinfo{author}{M.~Rose}, \bibinfo{author}{J.~Dunphy}, \bibinfo{author}{S.~Behler}, \bibinfo{author}{M.~Salmeron},
\newblock \bibinfo{title}{Adsorption and energetics of isolated co molecules on pd(111)},
\newblock \bibinfo{journal}{Surface Science} \bibinfo{volume}{453} (\bibinfo{year}{2000}) \bibinfo{pages}{25--31}.
\bibitem[{Toyoshima et~al.(2013)Toyoshima, Yoshida, Monya, Suzuki, Amemiya, Mase, Mun, and Kondoh}]{no_1}
\bibinfo{author}{R.~Toyoshima}, \bibinfo{author}{M.~Yoshida}, \bibinfo{author}{Y.~Monya}, \bibinfo{author}{K.~Suzuki}, \bibinfo{author}{K.~Amemiya}, \bibinfo{author}{K.~Mase}, \bibinfo{author}{B.~S. Mun}, \bibinfo{author}{H.~Kondoh},
\newblock \bibinfo{title}{Photoelectron spectroscopic study of co and no adsorption on pd(100) surface under ambient pressure conditions},
\newblock \bibinfo{journal}{Surface Science} \bibinfo{volume}{615} (\bibinfo{year}{2013}) \bibinfo{pages}{33--40}.
\bibitem[{Jia et~al.(2008)Jia, Yu, Wang, and Yongshan}]{no_2}
\bibinfo{author}{X.~Jia}, \bibinfo{author}{S.~Yu}, \bibinfo{author}{Z.~Wang}, \bibinfo{author}{m.~Yongshan},
\newblock \bibinfo{title}{Theoretical study of no adsorption/dissociation on pd (100) and (111) surfaces},
\newblock \bibinfo{journal}{Surface and Interface Analysis} \bibinfo{volume}{40} (\bibinfo{year}{2008}) \bibinfo{pages}{1350 -- 1355}.
\bibitem[{Jaworowski et~al.(2002)Jaworowski, Ásmundsson, Uvdal, and Sandell}]{no_3}
\bibinfo{author}{A.~Jaworowski}, \bibinfo{author}{R.~Ásmundsson}, \bibinfo{author}{P.~Uvdal}, \bibinfo{author}{A.~Sandell},
\newblock \bibinfo{title}{Determination of no adsorption sites on pd(100) using core level photoemission and low energy electron diffraction},
\newblock \bibinfo{journal}{Surface Science} \bibinfo{volume}{501} (\bibinfo{year}{2002}) \bibinfo{pages}{74--82}.
\bibitem[{Jung(2023)}]{gap_dataset}
\bibinfo{author}{H.~Jung}, \bibinfo{title}{{Local/Global minima adsorption structures on Rh(111)/(211)}}, \bibinfo{year}{2023}. \URLprefix \url{https://figshare.com/articles/dataset/Local_Global_minima_adsorption_structures_on_Rh_111_211_/23285156}. \DOIprefix\doi{10.6084/m9.figshare.23285156.v1}.
\bibitem[{Giannozzi et~al.(2017)Giannozzi, Andreussi, Brumme, Bunau, Nardelli, Calandra, Car, Cavazzoni, Ceresoli, Cococcioni et~al.}]{QS}
\bibinfo{author}{P.~Giannozzi}, \bibinfo{author}{O.~Andreussi}, \bibinfo{author}{T.~Brumme}, \bibinfo{author}{O.~Bunau}, \bibinfo{author}{M.~B. Nardelli}, \bibinfo{author}{M.~Calandra}, \bibinfo{author}{R.~Car}, \bibinfo{author}{C.~Cavazzoni}, \bibinfo{author}{D.~Ceresoli}, \bibinfo{author}{M.~Cococcioni}, et~al.,
\newblock \bibinfo{title}{Advanced capabilities for materials modelling with quantum espresso},
\newblock \bibinfo{journal}{Journal of physics: Condensed matter} \bibinfo{volume}{29} (\bibinfo{year}{2017}) \bibinfo{pages}{465901}.
\bibitem[{Wellendorff et~al.(2012)Wellendorff, Lundgaard, M\o{}gelh\o{}j, Petzold, Landis, N\o{}rskov, Bligaard, and Jacobsen}]{PhysRevB.85.235149_BElib}
\bibinfo{author}{J.~Wellendorff}, \bibinfo{author}{K.~T. Lundgaard}, \bibinfo{author}{A.~M\o{}gelh\o{}j}, \bibinfo{author}{V.~Petzold}, \bibinfo{author}{D.~D. Landis}, \bibinfo{author}{J.~K. N\o{}rskov}, \bibinfo{author}{T.~Bligaard}, \bibinfo{author}{K.~W. Jacobsen},
\newblock \bibinfo{title}{Density functionals for surface science: Exchange-correlation model development with bayesian error estimation},
\newblock \bibinfo{journal}{Phys. Rev. B} \bibinfo{volume}{85} (\bibinfo{year}{2012}) \bibinfo{pages}{235149}.
\bibitem[{Adhikari et~al.(2021)Adhikari, Nepal, Tang, and Ruzsinszky}]{C6H6_Ag111_hcp30}
\bibinfo{author}{S.~Adhikari}, \bibinfo{author}{N.~K. Nepal}, \bibinfo{author}{H.~Tang}, \bibinfo{author}{A.~Ruzsinszky},
\newblock \bibinfo{title}{Describing adsorption of benzene, thiophene, and xenon on coinage metals by using the zaremba–kohn theory-based model},
\newblock \bibinfo{journal}{The Journal of Chemical Physics} \bibinfo{volume}{154} (\bibinfo{year}{2021}) \bibinfo{pages}{124705}.
\bibitem[{Podryabinkin et~al.(2023)Podryabinkin, Garifullin, Shapeev, and Novikov}]{MLIP-3}
\bibinfo{author}{E.~Podryabinkin}, \bibinfo{author}{K.~Garifullin}, \bibinfo{author}{A.~Shapeev}, \bibinfo{author}{I.~Novikov},
\newblock \bibinfo{title}{Mlip-3: Active learning on atomic environments with moment tensor potentials},
\newblock \bibinfo{journal}{The Journal of Chemical Physics} \bibinfo{volume}{159} (\bibinfo{year}{2023}) \bibinfo{pages}{084112}.
\bibitem[{Feng et~al.(2024)Feng, Wang, and Pan}]{surf_rec}
\bibinfo{author}{J.~Feng}, \bibinfo{author}{X.~Wang}, \bibinfo{author}{H.~Pan},
\newblock \bibinfo{title}{In-situ reconstruction of catalyst in electrocatalysis},
\newblock \bibinfo{journal}{Advanced Materials} \bibinfo{volume}{36} (\bibinfo{year}{2024}) \bibinfo{pages}{2411688}.

\end{thebibliography}

\end{document}